\documentstyle[pre,aps,epsfig,floats,twocolumn]{revtex}
\begin{document}
\draft
\twocolumn[\hsize\textwidth\columnwidth\hsize\csname 
@twocolumnfalse\endcsname

\title{Percolation and depinning transitions in cut-and-paste models of adaptation}
\author{R. D'Hulst and G.J. Rodgers}
\address{Rene.DHulst@brunel.ac.uk and G.J.Rodgers@brunel.ac.uk}
\address{Department of Mathematical Sciences, Brunel University}
\address{Uxbridge, Middlesex, UB8 3PH, UK}
\date{\today}

\maketitle
\widetext
\begin{abstract}
We show that a cut-and-paste model to mimic a trial-and-error process of adaptation displays two pairs of percolation and depinning transitions, one for persistence and the other for efficiency. The percolation transition signals the onset of a property and the depinning transition, the growth of the same property. Despite its simplicity, the cut-and-paste model is qualitatively the same as the Minority Game. A majority cut-and-paste model is also introduced, to mimic the spread of a trend. When both models are iterated, the majority model reaches a frozen state while the minority model converges towards an alternate state. We show that a transition from the frozen to the alternate state occurs in the limit of a non-adaptive system.

\vspace{0.2cm}
\vspace{0.2cm}
\end{abstract}

]
\narrowtext

\section{Introduction}

The art of representing complex systems with simple models is based on abstracting towards an ideal system, keeping only the relevant features. Trying to keep the essential and discarding the superfluous. This approach has been successfully applied in physics, with the Ising model as paradigm of simplicity \cite{l}, and is now being exported to a variety of research fields, for instance with the growth of random networks \cite{krr}, the repetitions of ancestors in genealogical trees \cite{derrida} or the behaviour of flocks of birds \cite{vcbcs}. As an example, the Bak-Sneppen model \cite{bak-sneppen} allows one to simulate the dynamics of the natural co-evolution of many interacting species with a few lines of programming code, and the results compare well with empirical data. 

Another example is the Minority Game \cite{cz}, which represents the competition for scarce resources by reducing the dynamics to the search for minority between two groups. Developed in the framework of financial markets, it is usual to consider that this model represents agents having to guess whether buying or selling is the best choice for the next round of transactions. We present the Minority Game in greater detail in Sec. \ref{sec:the minority game}, and show that eventhough it appears to be very simple, it requires a certain level of complexity to represent many agents competing against one another. This complexity has been just enough to prevent the discovery of an exact solution to the model. In fact, all analytical approaches of the model \cite{cmz,gms,hjjh} depend at some point on rather intricate calculations. To avoid these complications, in this paper we analyse a macroscopic model based on the dynamics of the Minority Game, without modeling explicitly the microscopic dynamics of interacting agents \cite{dr}. 

This work has three basic motivations. First, microscopic quantities are rarely accessible in real life systems. Hence, it seems natural to investigate how microscopic models relate to some macroscopic dynamics. Second, the purpose of the Minority Game is oversimplified, guessing a minority group, but appears surprisingly out of reach of exact analytical solutions. We analyze here a much simpler model, which displays similar transitions and is exactly solvable. Third, we hope that, by introducing complexity step by step in our simple exactly solvable model, a better understanding of microscopic models will be achieved.

\section{The Minority Game}
\label{sec:the minority game}

The Minority Game is a generic model of competition for scarce resources, with applications to financial markets. A set of $N$ agents compete recursively to be in the minority group amongst buyers and sellers. They have to decide which action to choose between buying or selling at each time step. It is assumed that being in the minority group allows them to shop around and get a better deal. Each agent analyzes the outcome of the game for the $m$ previous time steps, the history of the system, and uses one of her $s$ strategies to make her next decision. She ranks her strategies according to their previous record of forecasting the minority group, regardless of whether these strategies where used or not. An agent always uses her strategy with the highest success rate for predictions.

\begin{figure}[h]
\centerline{\psfig{file=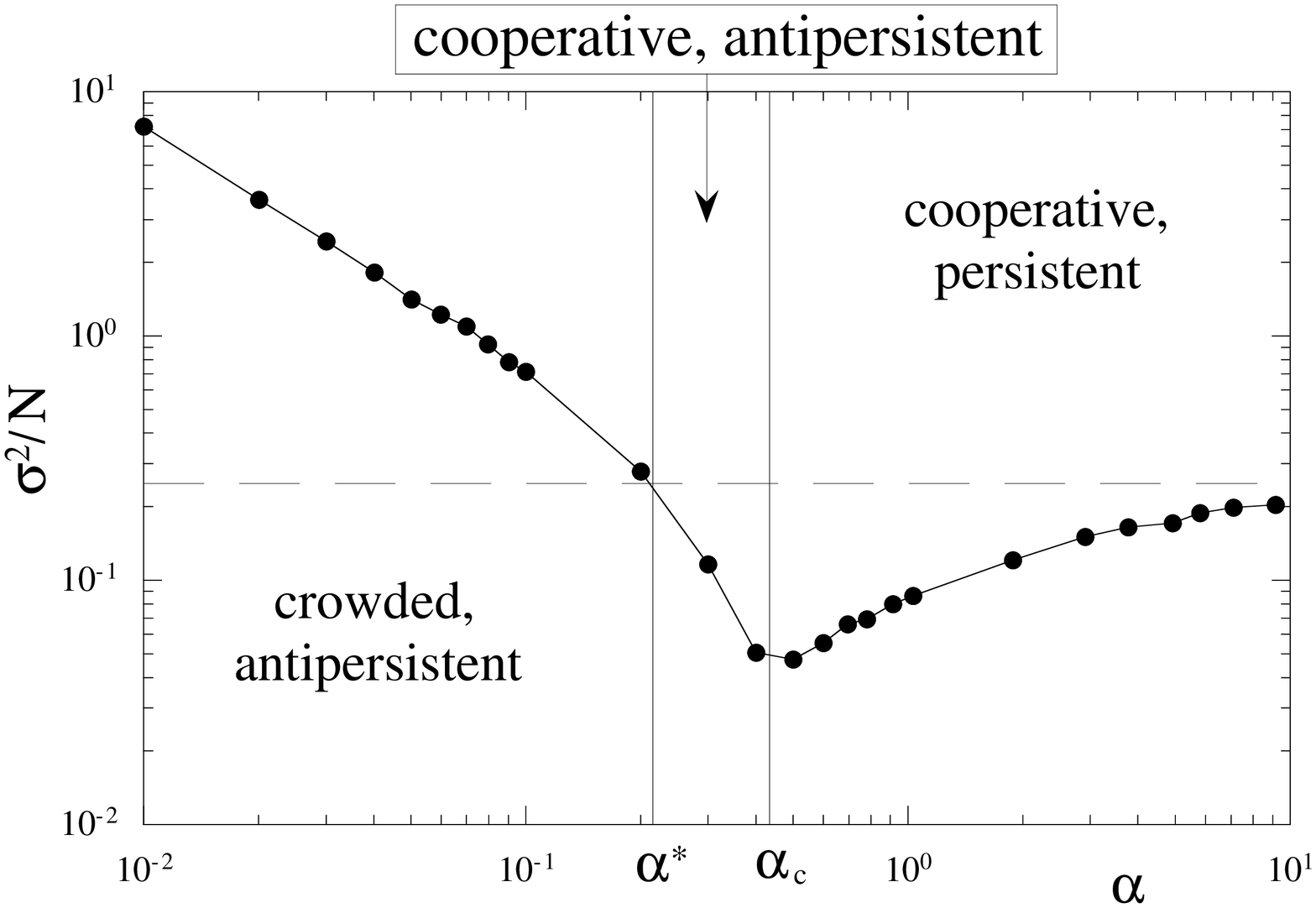,width=8.5cm}}
\caption{Variance $\sigma^2$ of the number of agents making one choice divided by the number of agents $N$, as a function of $\alpha$ for agents with two strategies at their disposal. The horizontal dashed lines refers to the expected result for random systems, $\sigma^2 /N = 1/4$.}
\label{fig:alp_intro}
\end{figure}

If it is assumed that the agents in the minority group are rewarded with a point and that the others get nothing, the system is most efficient whenever there are as many buyers as sellers. Here, efficiency would refer to the number of points given in any round of transactions, with $N/2$ as an upper limit. 
Hence, the deviation of the size of the two groups from $N/2$ is a direct measure of the lack of efficiency of the system or the lack of coordination between agents. There are many different measures for these deviations but the one most commonly used is the variance, defined as

\begin{equation}
\sigma^2 = \sum_{X=0}^N F (X) \left( X - \frac{N}{2}\right)^2
\end{equation}
where $F (X)$ is the probability of having $X$ buyers. If the agents were guessing at random between buying and selling, we would expect $\sigma^2_{rand} = N/4$. It is an important property of the Minority Game that $\sigma^2$ can either be greater than or less than $\sigma^2_{rand}$, depending on the particular choice of the model's three parameters, $(N,m,s)$. Remember that these are the number of agents, the length of the history and the number of strategies an agent has at her disposal, respectively. Using $\alpha = 2^m/N$, it is possible to make all the different values of $\sigma^2$ for one value of $s$ collapse on a single curve, as shown in Fig. \ref{fig:alp_intro} for $s=2$.
%fig1: variance 

In Fig. \ref{fig:alp_intro}, we distinguish three different `phases' according to the value of $\sigma^2$, crowded antipersistent for $\alpha < \alpha^{*}$, cooperative antipersistent for $\alpha^{*} < \alpha < \alpha_c$ and cooperative persistent for $\alpha > \alpha_c$. The crowded to cooperative transition at $\alpha^{*} \approx 0.21$ for $s=2$ corresponds to the transition from $\sigma^2$ greater than $\sigma^2_{rand}$ to $\sigma^2$ less than this value. In simple terms, it means that for $\alpha < \alpha^{*}$, the agents would be better off not adapting because they are doing worse than just guessing at random. For $\alpha> \alpha^{*}$, the adaptation process improves the global wealth of the system. The other special point, at $\alpha_c \approx 0.45$ for $s=2$, is at the minimum of $\sigma^2$. If $\sigma^2$ measures the system efficiency, it corresponds to the best the agents can achieve. However, it has been shown\cite{cmz} that the adaptive behaviour of the agents in the Minority Game is based on an analysis of the system persistence rather than its efficiency. To understand this notion of persistence, it suffices to remember that the Minority Game is based on guessing the minority choice. If for a particular history, selling remains the minority choice, the system is persistent for this history, while if systematically, buying and selling are alternatively the minority choice, the system is antipersistent. As we will show, the minimum in $\sigma^2$ does not correspond to the antipersistent-persistent transition. From now on, we will use the notation $\alpha_{ap}$ for the antipersistent-persistent transition and $\alpha_c$ for the minimum of the variance.

As we have seen, the Minority Game is based on a simple idea, guessing the minority choice, but it can be rather complex in its formulation, because it implies modeling the microscopic behaviour of $N$ agents. One can identify three main streams for current research; an analogy with spin glasses \cite{cmz}, challenged by a different dynamical formulation \cite{gms} and a crowd-anticrowd model \cite{hjjh}. The first two methods are conceptually attractive because of the clear analogy with methods borrowed from physics, and the nice physical picture of the antipersistent to persistent transition. The major restrictions are that Ref. \cite{cmz} fails to reproduce quantitatively the $\alpha < \alpha_{ap}$ behaviour of $\sigma^2$ and for both Refs. \cite{cmz,gms}, that the final `solution' is an implicit equation to be solved numerically. The latter approach by Johnson and co-workers \cite{hjjh} is interesting because of its own picture of the crowd-anticrowd interactions and more importantly, for the accurate reproduction of numerical simulations of the model. This approach has the limitation that the solution has equations involving summations that are not analytically tractable, and again need to be performed numerically to obtain explicit comparisons with the model. 

%fig2: explaning the transition
\setlength{\unitlength}{1pt}
\begin{figure}[h]
\begin{center}
\begin{picture}(240,300)(30,-60)
% lower boxes
\put(30,15){\dashbox{4}(240,0){}}
\put(268,25){$\frac{N}{2}$}
\put(36,-25){\framebox(31,80){}}
\put(50,47){0}
\put(73,-25){\framebox(31,80){}}
\put(87,47){1}
\put(116,-25){\framebox(31,80){}}
\put(130,47){0}
\put(153,-25){\framebox(31,80){}}
\put(167,47){1}
\put(196,-25){\framebox(31,80){}}
\put(210,47){0}
\put(233,-25){\framebox(31,80){}}
\put(247,47){1}
% filling of the lower boxes
\multiput(40,-25)(4,0){7}{\line(0,1){70}}
\put(36,45){\line(1,0){31}}
\multiput(77,-25)(4,0){7}{\line(0,1){10}}
\put(73,-15){\line(1,0){31}}
\multiput(120,-25)(4,0){7}{\line(0,1){45}}
\put(116,20){\line(1,0){31}}
\multiput(157,-25)(4,0){7}{\line(0,1){35}}
\put(153,10){\line(1,0){31}}
\multiput(200,-25)(4,0){7}{\line(0,1){35}}
\put(196,10){\line(1,0){31}}
\multiput(237,-25)(4,0){7}{\line(0,1){45}}
\put(233,20){\line(1,0){31}}
% upper boxes
\put(110,160){\dashbox{4}(80,0){}}
\put(200,160){$\frac{N}{2}$}
\put(116,120){\framebox(31,80){}}
\put(130,192){0}
\put(153,120){\framebox(31,80){}}
\put(167,192){1}
% filling of the upper boxes
\multiput(120,120)(4,0){7}{\line(0,1){20}}
\put(116,140){\line(1,0){31}}
\multiput(157,120)(4,0){7}{\line(0,1){60}}
\put(153,180){\line(1,0){31}}
% horizontal axis
\put(30,-45){\vector(1,0){240}}
\put(110,-55){$\alpha^{*}$}
\put(110,-45){\line(0,1){4}}
\put(150,-60){$\alpha$}
\put(190,-55){$\alpha_c$}
\put(190,-45){\line(0,1){4}}
% vertical axis
%\put(24,220){\vector(0,-1){190}}
%\put(0,128){time}
% labels for upper boxes
\put(117,225){When there is}
\put(116,210){no information:}
\put(30,185){The agents guess}
\put(30,170){at random}
\put(30,155){between 0 or 1.}
\put(30,140){Here, 0 wins.}
% labels for lower boxes
\put(65,95){The agents can adapt in three different}
\put(80,80){ways to the previous information.}
\put(60,65){Which way will be chosen, depends on $\alpha$:}
%\put(50,160){Here, 0 wins.}
\put(62,-35){(a)}
\put(146,-35){(b)}
\put(220,-35){(c)}
\end{picture}
\end{center}
\caption[Schematic explanation of the crowded to cooperative transition]{Schematic explanation for Fig. \ref{fig:alp_intro}. The different possible adaptations to the situation on the top of the figure are (a) crowded antipersisent, (b) cooperative antipersistent and (c) cooperative persistent.}
\label{fig:expla MG}
\end{figure}
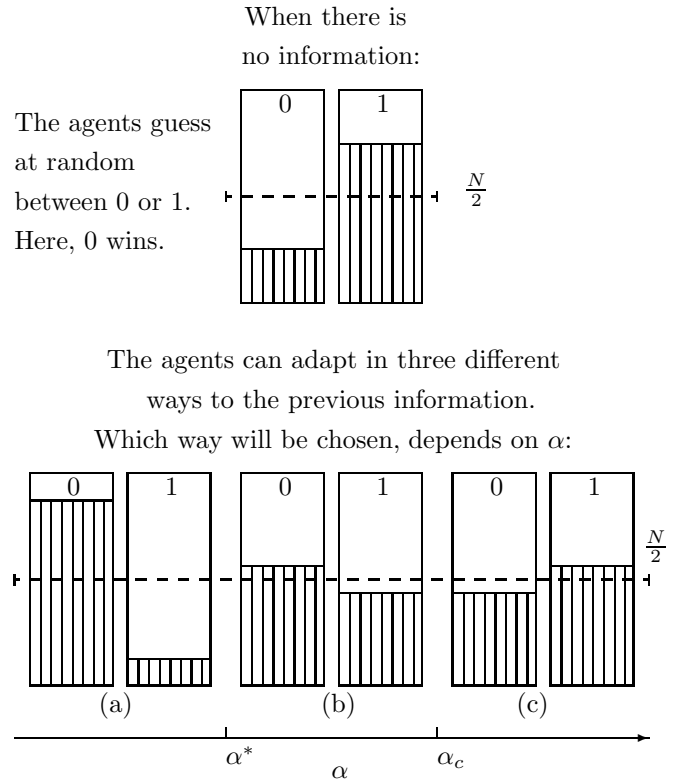
\setlength{\unitlength}{1cm}
In fact, it is possible to give a very basic picture to explain Fig. \ref{fig:alp_intro}. Consider the beginning of the game where agents are guessing at random between buying (0) or selling (1) because they don't have any information before playing. This corresponds to the higher part of Fig. \ref{fig:expla MG}, where we have schematically represented the two choices by two boxes, and the population in each box is represented by the dashed area. In the top of Fig. \ref{fig:expla MG}, we have represented an example where 0 is the winning choice. Agents can react in three different ways to this information. First, they can over-react and you will find most of them in box 0 at a later time, case $(a)$. This corresponds to what we called the crowded antipersistent phase, which appears for $\alpha < \alpha^{*}$. Second, another possible reaction is that agents over-react in the sense that the majority of them again pick choice 0, but less than just a random choice between 0 and 1. This is illustrated as case $(b)$, and it corresponds to the antipersistent, cooperative phase, with $\alpha^{*} < \alpha < \alpha_c$. Third and finally, agents react, but not enough of them are able to switch to 0, so that 0 remains the best choice for succeeding time steps, as illustrated in case $(c)$. This is a persistent, cooperative phase.

From the previous simple analysis, one sees that the important factor in the Minority Game is the degree of freedom left to the agents. If agents are free to adapt, they act very badly because they all react the same way, while if they have no freedom at all, they seem to act randomly, while in fact they just keep on doing exactly the same thing. Between these two extremes, there exists a balance where agents can be over-reacting or under-reacting, but still do better than just guessing at random. It is the object of this paper to consider a simple model to study this delicate balance between freedom and restriction of adaptation.

\section{The cut-and-paste model}

\begin{figure}[h]
\centerline{\psfig{file=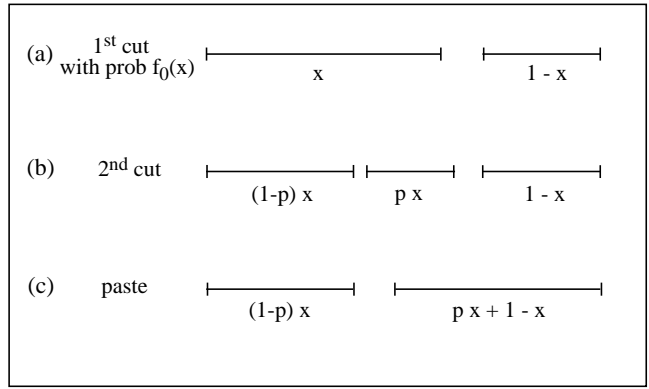,width=8.5cm}}
\caption{Schematic view of the cut-and-paste model}
\label{fig1:themodel}
\end{figure}

Our model works as follows. First, an interval of unit length is cut into two pieces of length $x$ and $1-x$ according to a distribution $f_0 (x)$, as shown in Fig. \ref{fig1:themodel}~(a). This first step is considered as a random guess for a value between 0 and 1. It is tacitly assumed that some microscopic dynamics are at work to produce this first cut, but we avoid modeling this process explicitly. It could correspond, for instance, to agents having to decide whether to buy or to sell a commodity without any previous information on its price. $x$ would be the fraction of buyers and $1-x$, the fraction of sellers. Then, a fraction $p$ of the bigger piece is cut and pasted to the smaller piece, as shown in Fig. \ref{fig1:themodel}~(b) and (c), respectively. This second step transforms the initial distribution $f_0 (x)$ into an adapted distribution $f_1 (x)$, in response to the first cut. We could say that $f_0 (x)$ is a distribution of random guesses while $f_1 (x)$ is a distribution of `educated guesses'. For the example of agents, after a first round of transactions, they become aware that there are either more buyers or more sellers, and react to this information, modifying the supply and demand equilibrium. The ideal reaction would result in equal size pieces, or the same number of buyers and sellers. The obvious problem is that there exists no global control on systems such as financial markets to synchronise the agents' decisions. We assume that the model is played several times without any correlations between different realizations, apart from the fact that we always start with the same distribution $f_0 (x)$.
%fig3: the cut-and-paste model

If we compare this with the Minority Game, it is easy to see that we have removed all the complexity of the original model by not modeling explicitly the microscopic dynamics. All the complexity of the Minority Game is hidden in a macroscopic parameter $p$. We will show in the next section that this allows us to solve the model exactly.

\section{Properties}
\label{sec:properties}

\begin{figure}[h]
\centerline{\psfig{file=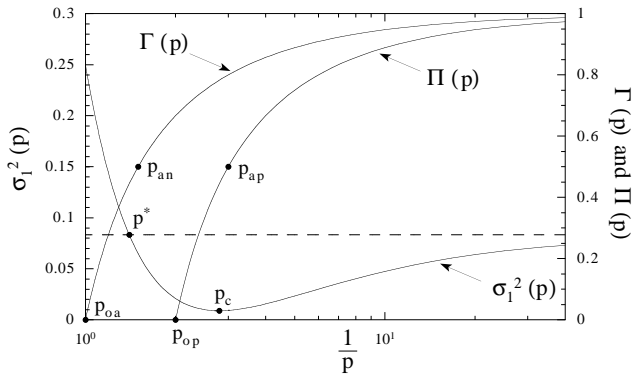,width=8.5cm}}
\caption{Variance $\sigma_1^2$ of $f_1$ as a function of $p$ (left scale), persistence probability $\Pi$ and improvment probability $\Gamma$ as functions of $p$ (both on the right scale). The initial distribution $f_0 (x)$ is uniform on $(0,1)$. The dots ($\bullet$) mark special values of $p$ considered in the text. The dashed horizontal line shows the variance of $f_0 (x)$.}
\label{fig2:uniform}
\end{figure}

For simplicity, we consider that $f_0 (x) = f_0 (1-x)$, that is, we incorporate a symmetry with respect to $1/2$, as in the Minority Game. Similarly, $f_1 (x) = f_1 (1-x)$, where $f_1 (x)$ is the size distribution after adaptation. This distribution is given by

\begin{eqnarray}
\nonumber
f_1 (x) &=& \int_0^{1/2} dy\ f_0 (y)\delta (y+p(1-y)-x)\\
 && + \int_{1/2}^1 dy\ f_0 (y) \delta (y(1-p) -x)
\label{eq:second-cut}
\end{eqnarray}
with a mean value $m_1 = 1/2$. The magnitude of the deviations from $m_1$ determine the efficiency of the system after adaptation. There are several quantities suited to measure these deviations but they all give qualitatively similar results, so that we choose the variance $\sigma^2_1$ of $f_1 (x)$, equal to

\begin{equation}
\sigma^2_1 = (1-p)^2 \sigma^2_0 + \frac{p(3p-2)}{4} + 2 p (1-p) K_0.
\label{eq:variance}
\end{equation}
$\sigma_0^2$ is the variance of $f_0(x)$, and we defined

\begin{equation}
K_0 \equiv \int_0^{1/2} x f_0 (x)\ dx.
\end{equation}
As in the Minority Game, two important values of $p$ are $p^{*}$ determined by $\sigma_1^2 (p^{*}) = \sigma_0^2$ and $p_c$ determined by

\begin{equation}
\left.\frac{\partial \sigma_1^2}{\partial p}\right|_{p_c} = 0.
\end{equation}
We obtain that 

\begin{equation}
p_{c} = \frac{4 \sigma^2_0 -  4 K_0 + 1}{4 \sigma^2_0 - 8 K_0 + 3}
\label{eq: critical p}
\end{equation}
and $p^{*} = 2 p_c$. This last relation should be general and in fact, in the Minority Game, $\alpha^{*} \approx \alpha_c / 2$. We will come back to this last relation in Sec. \ref{sec:comparison with the minority game}. To illustrate our results, we show $\sigma_1^2 (p)$ as a function of $1/p$ in Fig. \ref{fig2:uniform} for a uniform distribution $f_0 (x)$, which allows a visual comparison with Fig. \ref{fig:alp_intro}. For this particular choice, $p_c = 5/14$ and $p^{*} = 5/7$.
%Fig4: uniform distribution

However, remember that in the definition of the model, we assume that there are no correlations between different realizations of the model. Hence, $\sigma_1^2$ represents an average over several realizations of the model, but this is information that is not available to anyone seeing the game played only once. In other words, if we imagine an agent being part of the cut-and-paste model, he could not use $\sigma_1^2$ as a measure of global efficiency. This cannot be an interesting measure to predict if such an agent will decide afterwards if adapting was good or not. A more practical quantity is the probability $\Gamma (p)$ of improving on an initial guess. That is, $\Gamma (p)$ is the probability that the difference in length between the two pieces is smaller after adaptation. $\Gamma (p)$ is equal to

\begin{equation}
\Gamma (p) = 2 \int_0^{(1-p)/(2-p)} f_0 (x)\ dx,
\end{equation}
which is a monotonically decreasing function of $p$, equal to 1 when $p=0$ and 0 when $p=1$. Like $\sigma_1^2$, $\Gamma (p)$ is not a property directly available after one realisation of the model, but it gives a probability about the outcome of one such realisation. We show $\Gamma (p)$ as a function of $1/p$ for a uniform distribution $f_0 (x)$ in Fig. \ref{fig2:uniform}. 

$\Gamma (p)$ is a simple measure of global efficiency and two transitions can be associated with it, a percolation transition and a depinning transition. The percolation transition point is $p_{oa}= 1$, determined by $\Gamma (p_{oa}) = 0$. If we imagine that, after adapting, we want to measure if adaptation has improved coordination, $p_{oa}$ corresponds to the `onset of adaptation', meaning that for $p< p_{oa}$, it becomes possible that adaptation improves coordination. This transition is similar to a percolation transition \cite{stauffer}. Unfortunately, the $p> p_{oa}$ phase is not accessible as $p_{oa} = 1$. The order parameter of the transition is $\Gamma (p)$ which scales as 

\begin{equation}
\Gamma \sim (p_{oa}-p)^{\beta}
\end{equation}
for $p$ less than, but close to, $p_{oa}$. The value of the critical exponent $\beta$ depends on the analytical form of $f_0 (x)$. For instance, for a uniform distribution, $\beta = 1$, for $f (x) = 6 x (1 - x)$, $\beta = 2$ while for $f (x) = 2/\sqrt{x}$, $\beta = 1/2$. As in any percolation transition, there is no associated broken symmetry \cite{stauffer}. The other transition, the adaptive to non-adaptive transition is a depinning transition located in $p_{an}$ determined by $\Gamma (p_{an}) = 1/2$ \cite{barabasi}. This transition is associated to the fact that `intelligent agents' should be able to detect if they are doing better when adapting. For $p> p_{an}$, they would conclude that on average, the answer is no, while the answer is yes for $p< p_{an}$. A simple way to picture this transition is to consider a counter $A$ with the following dynamics. When the initial guess is improved, $A \to A + 1$, otherwise $A \to A - 1$, with $A\ge 0$. For $p> p_{an}$, $A$ stays close to 0, while for $p< p_{an}$, $A$ is `depinned' from 0 and goes away at a velocity $v_{an} = 1 - 2 \Gamma (p)$. This velocity, equal to 0 for $p>p_{an}$, is the order parameter of the transition \cite{barabasi}. For $p<p_{an}$, the symmetry between adapting or not adapting is broken. We show $\Gamma$ as a function of $1/p$ in Fig. \ref{fig2:uniform} for a uniform distribution $f_0 (x)$. With this choice, $p_{an} = 2/3$, while $p_{oa} = 1$ for any choice. 

A second important property of the system, after the efficiency, is the persistence. This property is described by the probability that the smaller piece remains the smaller piece after adaptation, that is, the persistence probability, $\Pi (p)$, is equal to
\begin{equation}
\Pi (p) = 2 \int_{0}^{(1-2p)/2(1-p)} f_0 (x)\ dx
\end{equation}
if $p<1/2$, zero otherwise. The persistence, like the efficiency, is characterized by a percolation and a depinning transition. The onset of persistence occurs when $p_{op} =1/2$, determined by $\Pi (p_{op}) = 0$. This is a percolation-like transition \cite{stauffer}, with $\Pi (p)$ as order parameter and no associated broken symmetry. The critical exponent $\beta$ determined by $\Pi \sim (p_{op} - p)^{\beta}$, has the same value for this transition and the transition in efficiency. For $p>1/2$, we define $\xi \equiv |1/2 - x|^{-1}$, where $x$ is the smallest piece than can be obtained after adapting. $\xi$ compares to the largest cluster size below the percolation threshold or to a correlation length. We find that 

\begin{equation}
\xi \sim (p - p_{op})^{-\nu}
\end{equation}
with $\nu = 1$, as in percolation \cite{stauffer}. The corresponding depinning transition is located at $p_{ap}$ determined by $\Pi (p_{ap}) = 1/2$. As for the depinning transition, it is best understood if we keep a record of previous realisations, with a parameter $B$ counting the asymmetry between the left and the right piece. When the left piece is the smaller or the larger piece before and after adapting, $B \to B + 1$, while $B \to B - 1$ otherwise, with the added condition that $B \ge 0$. For $p> p_{op}$, $B$ stays close to 0, while for $p< p_{ap}$, $B$ moves away from zero at a velocity $v_{ap} = 1 - 2 \Pi (p)$, $p = p_{ap}$ marking the depinning transition. This transition corresponds to a breaking of the left-right symmetry. In the Minority Game, the depinning transition in persistence has been identified using an analogy with spin glasses \cite{cmz}. For a uniform distribution, $p_{ap} = 1/3$. We show $\Pi (p)$ for a uniform distribution in Fig. \ref{fig2:uniform}.

\begin{table}[h]
\centering
\begin{tabular}{ccc}
%\hline
value of $p$ & corresponding property & example \\
\hline
$p_{ap}$ & antipersistence,persistence & 1/3 \\ 
$p_c$ & minimum of $\sigma_1^2$ &  5/14 \\ 
$p_{op} = 1/2$ & onset of persistence & 1/2 \\
$p_{an}$ & adaptive,non-adaptive & 2/3 \\
$p^{*}$ & crowded-cooperative &  5/7 \\ 
$p_{oa} = 1$ & onset of adaptation & 1 \\
%\hline
\end{tabular}
\caption{Particular values of $p$ (column 1), their meaning (column 2) and their values for a uniform distribution $f_0 (x)$ (column 3).}
\label{table:special p}
\end{table}
In Table \ref{table:special p}, we summarise the six different values of $p$ considered here, their meaning and their values for a uniform distribution $f_0 (x)$. From this table, we can appreciate the simple relation between the two properties of the model, persistence and efficiency. In fact, as explained in Ref. \cite{dr}, similar properties are characterized by $\Gamma (p_{eff}) = \Pi (p_{pers})$, which implies $p_{eff} = 2 p_{pers}$. For instance, the onsets are characterized by $\Pi (p_{op}) = \Gamma (p_{oa}) = 0$. From the previous relation, one obtains the relations $p_{oa} = 2 p_{op}$ and $p_{an} = 2 p_{ap}$. Hence, a characteristic property of persistence in the system requires twice as much reaction from the system to have the same effect on the system efficiency \cite{dr}.

It is important to realize that in Table \ref{table:special p}, the particular values of $p$ do not coincide with each other. A consequence of this fact is that we do not expect in general to have the antipersistence-persistence transition point that coincides with the minimum of the variance. For the Minority Game, the former has been found at $\alpha_{ap} \approx 0.34$, while we mentioned that the latter is at $\alpha_c \approx 0.45$, for $s=2$ strategies per agent. Moreover, it is easy to show that using other measures of the deviations with respect to 1/2, new distinct values of $p$ would be obtained. For instance, if the mean-absolute-deviation from 1/2, $E_{abs}$, is measured instead of $\sigma_1^2$, the minimum of $E_{abs}$ is not at $p_c$ and its random value is not at $p^{*}$. This fact can have a profound impact on the dynamics of any model which is based on a restricted adaptation aiming to optimize the efficiency. For instance, in the Minority Game, the agents are trying to minimize the information in the system, which is equivalent to saying that they analyze the persistence in the system, rather than the efficiency, as was already noted in \cite{cmz}.

%majority model:
\section{A majority cut-and-paste model}

The Minority Game is based on the assumption that agents are competing to guess the choice of the minority. In Ref. \cite{deCara}, a model where agents are competing to guess the choice of the majority was introduced. It was shown that the dynamics seem not as trivial as expected, with the agents being not very effective at acting in a coordinated manner. Similarly, in the majority cut-and-paste model, a part $p$ of the smaller piece is added to the larger one. If the minority cut-and-paste model is reminiscent of trial-and-error adaptation, the majority version models the spread of a trend. It identifies for instance to the choice of driving on the right or on the left, which is random initially and spreads by imitation before being enforced by law. Using the notations of the previous sections,

\begin{eqnarray}
\nonumber
f_1 (x) &=& \int_0^{1/2} dy\ f_0 (y)\delta (y(1-p)-x)\\
&+& \int_{1/2}^1 dy\ f_0 (y) \delta (y+p(1-y) -x)
\end{eqnarray}
which is equivalent to 

\begin{equation}
f_1 (x) = \frac{1}{1 - p} f_0 \left( \frac{x}{1-p}\right)
\end{equation}
for $0\le x \le (1-p)/2$,

\begin{equation}
f_1 (x) = \frac{1}{1 - p} f_0 \left( \frac{x - p}{1-p}\right)
\end{equation}
for $(1+p)/2\le x \le 1$ and $f_1 (x) = 0$ for $(1-p)/2\le x \le (1+p)/2$. This distribution is bimodal with a gap where $f_1 (x) = 0$ for $(1-p)/2\le x \le (1+p)/2$. As before, we suppose that $f_0 (x)$ is symmetric around $x = 1/2$, so that $f_1 (x)$ also shares this property. As explained in \cite{deCara}, the variance of $f_1 (x)$ is not an adequate measure for the system efficiency, which is better determined by the deviation from perfect coordination. Perfect coordination is defined by the longer piece being equal to $1$ or, in other words, the disappearance of the smaller piece. The quantity we consider is the mean deviation from $1$ of the longer piece, defined by

\begin{eqnarray}
\label{eq:mean-deviation majority, definition}
\mu^2_1 &\equiv& 2 \int_{1/2}^1 (1 - x)^2 f_1 (x) dx\\
&=& 2(1 - p)^2 \int_{1/2}^1 (1 - x)^2 f_0 (x) dx
\label{eq:mean-deviation majority}
\end{eqnarray}
which is monotonic in $p$, as expected. 

Of course, the efficiency in the model is now described by the probability of having the longer piece closer to $1$ after adaptation, while the persistence is not really a relevant property of the model. Nevertheless, if we consider these properties, the model is characterized by the fact that all the previously considered transition points are in $p=0$.

\section{Iterated models}
\label{sec:iterated models}

\subsection{The majority cut-and-paste model}

We imagine that the majority cut-and-paste model is applied recursively, starting with a distribution $f_0 (x)$, adapting to $f_1 (x)$ which becomes an `educated guess', creating $f_2 (x)$, which becomes the new educated guess, and so on. That is, after an initial random cut according to a trial distribution $f_0 (x)$, a fraction $p$ of the smaller piece is added to the larger one, giving a size distribution $f_1 (x)$. Then, a fraction $p$ of the smaller piece is added to the smaller one, creating a size distribution $f_2 (x)$, and so on $n$ times. The size distribution after $i$ different cuts, $f_i (x)$ obeys the recurrence relation

\begin{eqnarray}
\nonumber
f_i (x) &=& \int_0^{1/2} dy\ f_{i-1} (y)\delta (y(1-p)-x)\\
&+& \int_{1/2}^1 dy\ f_{i-1} (y) \delta (y+p(1-y) -x).
\label{eq:h of f - majority}
\end{eqnarray}
We assumed that $p$ is independent of the number of adaptations. The first moment of each distribution, $m_i$, is equal to $m_i = 1/2$, if we assume that $f_0 (x) = f_0 (1 - x)$, as before. 

%Fig5: opening of a gap
\begin{figure}[h]
\centerline{\psfig{file=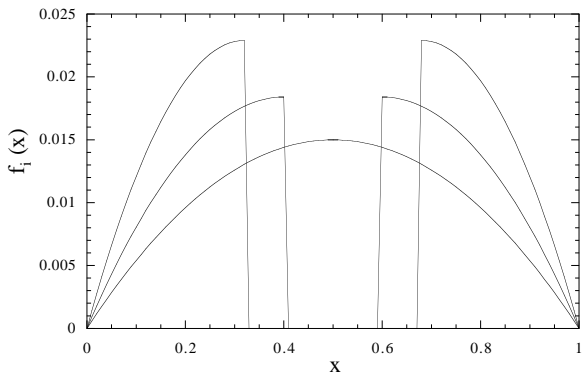,width=8.5cm}}
\caption{$f_i (x)$ for the majority model for $f_0 (x) = 6 x(1-x)$, $p=0.2$ and $i=0$, 1 and 2. A gap of length given by Eq. (\ref{eq:gap}) is opening in the middle as $i$ increases.}
\label{fig:gap}
\end{figure}
It is easy to show that the above recurrence relation induces the opening of a gap $G$ equal to 

\begin{equation}
G = 1 - (1-p)^i,
\label{eq:gap}
\end{equation}
where $f_i (x)$ is equal to 0, as illustrated in Fig. \ref{fig:gap}. This gap is centred around $x = 1/2$. Hence, the limit 

\begin{equation}
\lim_{i\to \infty} f_i (x) = \frac{\delta (x) + \delta (1 - x)}{2}
\end{equation}
for the majority model. Also, the mean deviation of $f_i (x)$ from $1$, $\mu_i^2$, defined like Eq. (\ref{eq:mean-deviation majority, definition}), is equal to

\begin{equation}
\mu^2_i = (1-p)^{2i} \mu^2_0,
\end{equation}
using Eq. (\ref{eq:mean-deviation majority}) iteratively. In fact, this conclusion can be generalized to the deviations from 1 of any order, with

\begin{eqnarray}
\mu^k_i &\equiv& 2 \int_{1/2}^1 (1 - x)^k f_i (x) dx\\
&=& (1 - p)^{ki} \mu^k_0. 
\end{eqnarray} 

\subsection{The minority cut-and-paste model}

The iterated minority cut-and-paste model is less trivial, even if the recurrence relation for the size distribution after $i$ cuts

\begin{eqnarray}
\nonumber
f_i (x) &=& \int_0^{1/2} dy\ f_{i-1} (y)\delta (y+p (1-y)-x)\\
&+& \int_{1/2}^1 dy\ f_{i-1} (y) \delta (y(1-p) -x),
\label{eq:h of f - minority}
\end{eqnarray}
looks very similar to the majority model. For the sake of clarity, the previous equation states that after an initial cut, according to $f_0 (x)$, a fraction $p$ of the larger piece is cut and pasted to the smaller piece, giving a distribution $f_1 (x)$; then a fraction $p$ of the larger piece is cut and pasted to the smaller piece, giving a distribution $f_2 (x)$; and so on $n$ times. The effect of the transformation depends on the value of $p$ considered, which determines the intervals where $f_i (x)$ is non zero. For $p=0$, $f_{i-1} (x)$ is simply not modified. For $0 < p \le 1/3$, $f_i (x)$ is non-zero for $x\in (p,1-p)$, that is, the initial function is modified and compressed on a narrower interval. For $1/3\le p \le 1/2$, the interval where $f_i (x)\not = 0$ increases with $p$, spanning from $(1-p)/2$ to $(1+p)/2$. Finally, for $p>1/2$, $f_i (x)$ is non-zero in two disconnected intervals, each interval being of size $(1-p)/2$. As a summary, for $p\le 1/2$, $f_{i-1} (x)$ is modified and squeezed in a smaller interval, while for $p>1/2$, $f_{i-1} (x)$ is modified and split into two disconnected intervals. This process is completely independent of the starting function $f_0 (x)$.

Starting with $p\le 1/2$, let us consider separately the cases when the smaller piece is of length less than $(1-2p)/2(1-p)$ and when it is longer than that. For the latter, one obtains the correspondence

\begin{equation}
\left( \frac{1 - 2p}{2 (1-p)} , \frac{1}{2}\right) \to \left( \frac{1 - p}{2} , \frac{1}{2}\right)
\end{equation}  
for the length of the smaller piece. That is, if the smaller piece is included in the interval on the left of the previous equation after $i$ cuts, it will be in the interval on the right after the $(i+1)^{\hbox{\small th}}$ cut. The interval on the left corresponds to the antipersistent points and, as it includes the interval on the right for $p\in (0,1/2)$, as soon as the system enters an antipersistent phase, it stays antipersistent. Now, the length of a piece that has been the smaller piece for $i$ successive adaptations, $x_i^{small}$, is equal to

\begin{equation}
x_i^{small} = 1 - (1-p)^i (1 - x^{small}_0).
\end{equation}
Whenever this is greater than $(1-2p)/2(1-p)$, the system becomes antipersistent and, as we have just shown, it stays antipersistent. So, \emph{the antipersistent phase is an attractor of the dynamics}. After $i$ cuts, all pieces that were of length greater than

\begin{equation}
x_{low} (i) = 1 - \frac{1}{2 (1-p)^{i+1}}
\label{eq: x low for antipersistence}
\end{equation}
have entered the antipersistent attractor. For a given value of $p$, the system has entered the attractor after

\begin{equation}
i_{att} = \frac{\ln \frac{1}{2}}{\ln (1-p)} - 1
\end{equation}
adaptations for all initial conditions. An alternative way of expressing this property is to look at the persistence probability, which is equal to

\begin{equation}
\Pi_i (p) = 2 \int_0^{x_{low}} f(x) dx
\label{eq:persistence after i cuts}
\end{equation}
after $i$ adaptive cuts for $i < i_{att}$, and $\Pi_i (p) = 0$ otherwise. As $x_{low}Ê(i) \to 0$ when $i = i_{att}$, $\Pi_i (p)$ is identically equal to zero when $i\ge i_{att}$. That is, the system becomes completely antipersistent.

The main conclusion of our analysis is that the system can be persistent for a transient period, depending on the initial conditions, but that it will eventually enter an antipersistent phase if it is allowed to evolve ad infinitum. This is the phase we now consider for any value of $p$. That is, the left piece is the smaller and the larger alternately. The length of the larger piece after $i$ adaptations is equal to

\begin{equation}
x_i^{large} = p \sum_{j=0}^{[(i-1)/2]} (1-p)^{2j} + (1-p)^i x_0
\end{equation}
where $x_0$ is the size of this same piece after the first cut. We use the notation $[(i-1)/2] = (i-1)/2$ for $i$ odd and $i/2-1$ for $i$ even. The length of the larger piece converges towards $\lim_{i\to \infty} x_i^{large} = 1/(2-p)$, independently of the initial conditions. One could say that the system becomes fully efficient only for a quasistatic adaptation, meaning that the larger and the smaller pieces become equal only when $p \to 0^{+}$. The size difference between the larger and smaller piece after $i$ adaptations, $\Delta x_i$, is equal to

\begin{equation}
\Delta x_i = \frac{p}{2-p} (1 - (p-1)^i) + 2(p-1)^i \left( x^{large}_0 - \frac{1}{2} \right),
\end{equation}
which converges towards $\lim_{i\to \infty} \Delta x_i = p/(2-p)$. Hence, we have

\begin{equation}
\lim_{i\to \infty} f_i (x) = \frac{\delta \left( x - \frac{1}{2-p} \right) + \delta \left( x - \frac{(1-p)}{2-p}\right) }{2}.
\end{equation}
The variance of $f_i (x)$ is given by

\begin{eqnarray}
\nonumber
\sigma_i^2 &=& (1-p)^{2i} \sigma^2_0 + \frac{p^2 (1 - (p-1)^i)^2}{4 (2-p)^2}\\
&+& \frac{p (p-1)^i (1 - (p-1)^i)}{2-p}Ê\left( \frac{1}{2} - 2 K_0\right).
\end{eqnarray}
For large $i$ and $p\not = 0$, $\sigma_i^2 \simeq p^2/4(2-p)^2$, independent of $f_0 (x)$. 

A simple way to mix both the minority and majority cut-and-paste models is to let $p$ take negative values. Suppose we consider the minority cut-and-paste model. Having $p< 0$ simply means that the size of the larger piece increases, which is equivalent to the majority model. One should however be careful because, still considering the minority model, $p$ cannot be less than $1 - 1/x$, where $x$ is the length of the largest piece. Hence, such an implementation is not very practical as it requires different boundaries for $p$ for different realisations. Nevertheless, it shows that $p = 0$ marks the transition from minority to majority. When the models are iterated, $p = 0$ marks the transition point between a frozen state and an alternate state. This is equivalent to a transition from a broken symmetry to a temporarily-broken symmetry. These two different states were also found in the framework of collective motion. It was shown that most models are associated to a broken symmetry \cite{vcbcs}, while one of them displays an alternate state, with a temporarily-broken symmetry \cite{oe}. However, no transition from one to the other was seen before.

\section{Comparison with the Minority Game}
\label{sec:comparison with the minority game}

Strictly speaking, the basic cut-and-paste model only compares with the Minority Game when $m=0$. This is related to the dynamics of the game which make the agents diffuse on the edges of a $2^m$ dimensional hypercube instead of a line. Each edge corresponds to one of the $2^{2^m}$ different strategies. When a game is played, this is equivalent to cutting the cube according to one of its $2^m$ median planes and counting the number of agents on each side of the plane. The agents on the minority side are winning, and the others react by trying to diffuse from where they are to the minority side. One component of the complexity of the diffusion process is that agents are only given $s$ strategies chosen at random at the beginning of the game, so that they are jumping between uncorrelated locations. The other component is that they do not know the direction of the next median plane. Hence, they are adapting for past information, not trying to forecast the next output of the game. Of course, we do not want to consider the whole complexity of the original model, our aim being to introduce something much simpler. We want to provide a qualitative comparison between the cut-and-paste model and the Minority Game.

In the Minority Game, an agent is able to adapt for a particular history if, among her $s$ strategies, she has both buying and selling as decisions for this particular history. With an initially random attribution of the strategies, the average adaptation potential of the whole system, $\overline{p}$, is given by

\begin{equation}
\overline{p} = 1 - \frac{1}{2^{s-1}}.
\label{eq:potential p of the MG} 
\end{equation}  
This value means that an average of $\overline{p}N$ agents are effectively able to make a choice for a history, while $(1-\overline{p})N$ agents have strategies that all make the same predictions for this history. Unfortunately, the agents do not know which history will come next and even if they are able to adapt, they cannot synchronise perfectly. The random guess distribution, $f_0 (X)$, is given by

\begin{equation}
f_0 (X) = \frac{N!}{2^N X! (N - X)!},
\end{equation}
where $X$ can only take integer values in $(0, N)$. From this distribution, one has $\sigma^2_0 = N/4$ and 

\begin{equation}
K_0 = \frac{N}{4} - \frac{N!}{2^{N+1} \left( \frac{N-1}{2}\right)!^2}. 
\end{equation}
The variance is given by

\begin{equation}
\sigma^2_1 = \frac{(1-p)^2N}{4} + \frac{N^2p^2}{4} - N p (1-p) \frac{N!}{2^{N} \left( \frac{N-1}{2}\right)!^2}.
\label{eq:variance for cut-and-paste}
\end{equation}
We use this last relation to compare the cut-and-paste model to the Minority Game.

According to the crowd-anticrowd theory of the Minority Game \cite{hjjh}, $v/2 < \sigma^2/N < v$ for $\alpha < \alpha_c$, where

\begin{equation}
v = \frac{N}{3.2^{m+2}} \left( 1 - \frac{1}{2^{2(m+1)}} \right).
\end{equation}
For definiteness, we take $\sigma^2/N \approx 3v/4$. The limit $\alpha < \alpha_c$ is similar to the $p\to 1$ limit. From Eq. (\ref{eq:variance for cut-and-paste}), we obtain that

\begin{equation}
p_{low} = \frac{1}{2^{\frac{m+1}{2}}} \left( 1 - \frac{1}{2^{2(m+1)}}Ê\right)
\end{equation}
when $p \approx 1$. The subscript `low' refers to $\alpha < \alpha_c$. In contrast, for  $\alpha > \alpha_c$,
\begin{equation}
\frac{\sigma^2}{N} = \frac{1}{4} \left( 1 - \frac{N}{2^{m+1}}\right)
\end{equation}
according to the crowd-anticrowd theory \cite{hjjh}. Compared with Eq. (\ref{eq:variance for cut-and-paste}) in the limit $p \to 0$, we obtain

\begin{equation}
p_{high} = \frac{1}{8\alpha} \left( \frac{1}{2} + \frac{N!}{2^N (\frac{N-1}{2})!^2} \right)^{-1}.
\label{eq:p_high}
\end{equation}
Similarly, the subscript `high' refers to $\alpha > \alpha_c$.

\begin{figure}[h]
\centerline{\psfig{file=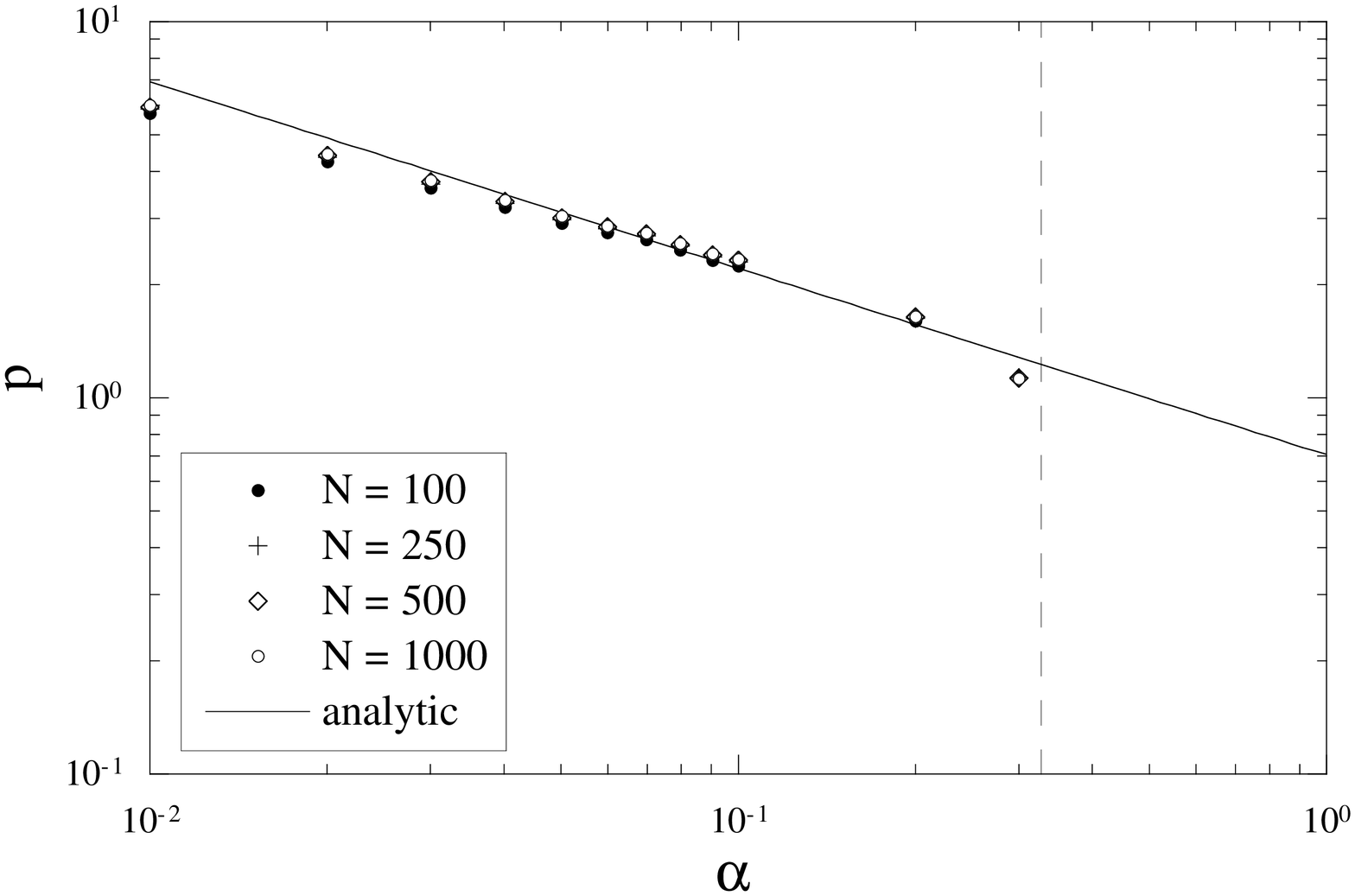,width=8.5cm}}
\centerline{\psfig{file=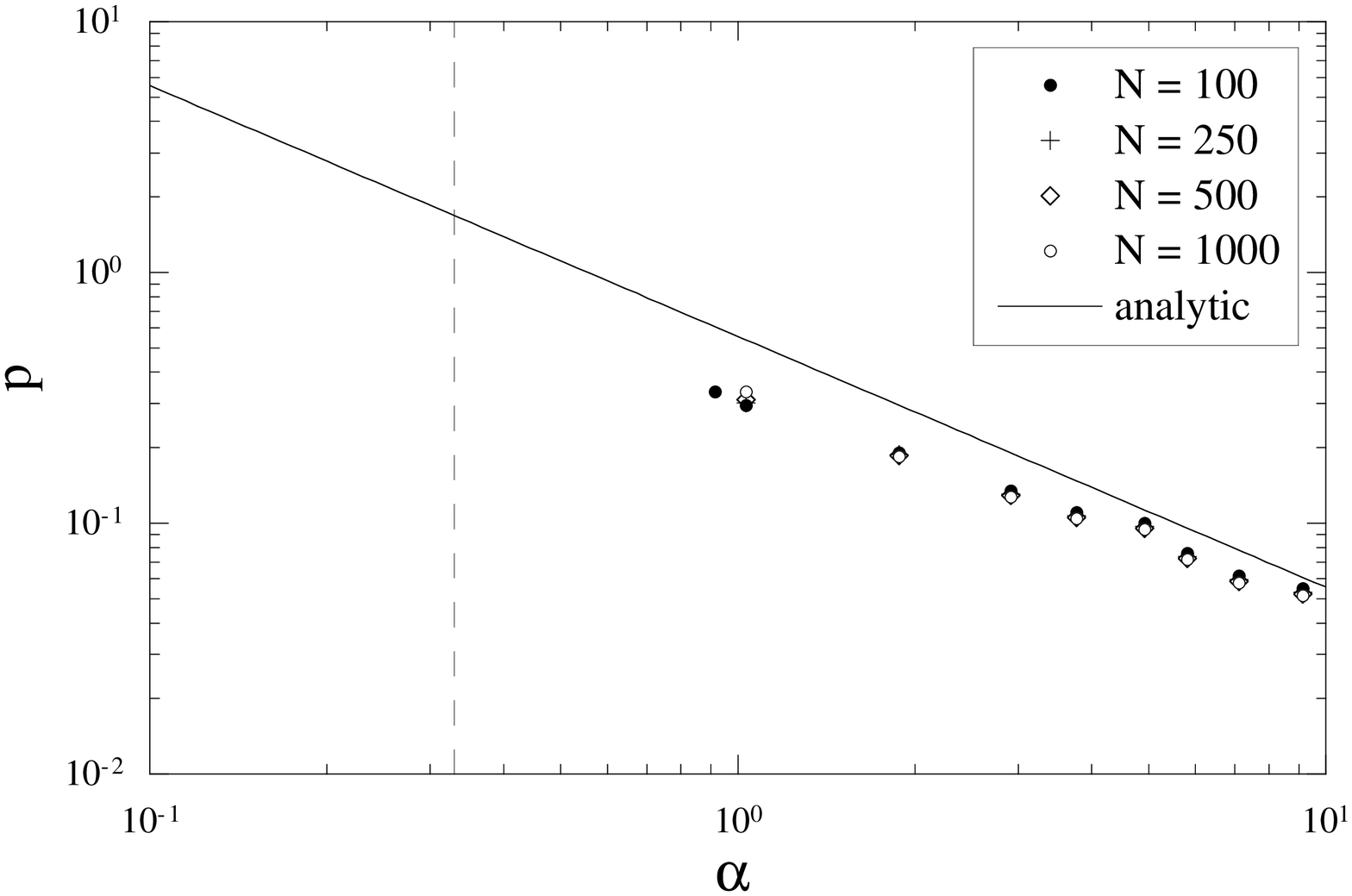,width=8.5cm}}
\caption{Variation of the fraction $p$ of the population that is adapting as a function of $\alpha$ in the Minority Game. For $\alpha < \alpha_c$ (top), we present $p\sqrt{N}$ as a function of $\alpha$ for $N= 100$ ($\bullet$), 250 (+), 500 ($\diamond$) and 1000 ($\circ$) while for $\alpha > \alpha_c$ (bottom), we present $p F(N)$ as a function of $\alpha$ for the same values of $N$. The factor $F (N)$ is given in Eq. (\ref{eq:p_high}). The continuous lines are the corresponding analytical expressions and the vertical dashed line shows the transition point $\alpha = \alpha_{ap}$.}
\label{fig:p_alpha}
\end{figure}

We can check the previous analytical results by finding numerically the relation between $p$ and $\alpha$. In fact, numerical simulations of the Minority Game give $\sigma^2$ as a function of $\alpha$ and solving numerically Eq. (\ref{eq:variance for cut-and-paste}) with respect to $p$, we obtain $p$ as a function of $\alpha$. The function $p (\alpha)$ is shown in Fig. \ref{fig:p_alpha}.
%fig6: p of alpha 
From this figure, one can appreciate that the predicted scalings are in good agreement with the numerical simulations. We have to stress that for $N \ge 8$, there are values of $\sigma^2 /N$ around $\alpha_c$ which are too small to be obtained for any value of $p$. That is, the minimum of $\sigma^2$ in the Minority Game is smaller than the minimum of $\sigma_1^2$ of the cut-and-paste model when $N \ge 8$. This limitation of the cut-and-paste model is related to the fact that we only analyzed a model where a constant ratio $p$ of loosers are adapting whatever the outcome of the initial random cut is. Further work could concentrate on a more accurate relation between $p$ and the parameters of the Minority Game using the history distribution for instance \cite{dr2}. Letting $p$ be chosen from a distribution could also be interesting.

\section{Conclusions}

In summary, we have presented a cut-and-paste model to mimic a trial-and-error process of adaptation. The model depends on only one parameter, $p$, that encodes the reactibility of the system to previous information. We show that the model displays two properties, persistence and efficiency. A pair of transitions is associated with each property. A pair is composed of a percolation transition, corresponding to the onset of a property, and a depinning transition, corresponding to the growth of the same property. We identified the critical exponents for these transitions, showing that both transitions share the same set of exponents. The exponent of the order parameter for the percolation transition depends on the initial conditions. Using a very simple analogy, we are able to propose a qualitative comparison with the Minority Game, keeping a more accurate comparison for further research. A majority cut-and-paste model is also introduced to model the spread of a trend. We show that all the transition points are at $p=0$. Both models are iterated, showing that the majority model reaches a frozen state, while the minority model converges towards an alternate steady state. The transition from one state to the other is achieved for $p=0$, that is, in the limit of non-adaptive systems. Our work should be relevant to the study of adaptive systems and particularly the Minority Game. In fact, we argue that the original Minority Game is itself an intrincate formulation of this simple mechanism, where the parameter $p$ is a complex quantity depending on the history given to the agents and the strategy space. This paper is a contribution towards the understanding of this complexity.

One of us (GJR) would like to thank The Leverhulme Trust for financial support.

\end{document}